# A physics-constrained machine learning method for mapping gapless land surface temperature

Jun Ma[a], Huanfeng Shen[a*], Menghui Jiang[a], Liupeng Lin[a], Chunlei Meng[b], Chao Zeng[a], Huifang Li[a], Penghai Wu[c]

[a] School of Resource and Environmental Sciences, Wuhan University, Wuhan 430079, China
[b] Institute of Urban Meteorology, China Meteorological Administration, Beijing 100089, China
[c] Information Materials and Intelligent Sensing Laboratory of Anhui Province, Anhui University, Hefei 230601, China

*Corresponding author.
*E-mail address:* shenhf@whu.edu.cn (H. Shen)

**Abstract:** More accurate, spatio-temporally, and physically consistent land surface temperature (LST) estimation has been a main interest in Earth system research. Developing physics-driven mechanism models and data-driven machine learning (ML) models are two major paradigms for gapless LST estimation, which have their respective advantages and disadvantages. In this paper, a physics-constrained ML model, which combines the strengths in the mechanism model and ML model, is proposed to generate gapless LST with physical meanings and high accuracy. The hybrid model employs ML as the primary architecture, under which the input variable physical constraints are incorporated to enhance the interpretability and extrapolation ability of the model. Specifically, the light gradient-boosting machine (LGBM) model, which uses only remote sensing data as input, serves as the pure ML model. Physical constraints (PCs) are coupled by further incorporating key Community Land Model (CLM) forcing data (cause) and CLM simulation data (effect) as inputs into the LGBM model. This integration forms the PC-LGBM model, which incorporates surface energy balance (SEB) constraints underlying the data in CLM-LST modeling within a biophysical framework. Results indicate that, for the sample-based validation (space-based validation), the root-mean-square error (RMSE) = 1.23–2.03 K (1.56–2.62 K) and the Pearson correlation coefficient (R) = 0.99 (0.98–0.99). Validation with four independent ground measurements shows that the generated clear-sky LST is comparable, or even better than the original Moderate Resolution Imaging Spectroradiometer- (MODIS) LST in specific scenarios. The generated LST also presents a high accuracy (RMSE = 2.91–3.66 K and R = 0.97–0.98) under cloudy-sky conditions. Compared with a pure physical method and pure ML methods, the PC-LGBM model improves the prediction accuracy and physical interpretability of LST. It also demonstrates a good extrapolation ability for the responses to extreme weather cases, suggesting that the PC-LGBM model enables not only empirical learning from data but also rationally derived from theory. Compared with other popular ML methods and sophisticated all-weather products, the PC-LGBM model delivers a superior validation accuracy and image quality. The proposed method represents an innovative way to map accurate and physically interpretable gapless LST, and could provide insights to accelerate knowledge discovery in land surface processes and data mining in geographical parameter estimation.

## 1. Introduction

Land surface temperature (LST) is a crucial parameter in land-atmosphere interactions, reflecting the surface energy balance (SEB) and fluxes exchange and is widely used in multi-disciplinary research, such as climatology (Hansen et al. 2010), urbanology (Shen et al. 2016), agriculture (Karnieli et al. 2010), ecology (Connors et al. 2013) and hydrology (Anderson et al. 2012). Accordingly, LST has been identified as one of the most essential climate variables (ECVs) by the Global Climate Observing System (GCOS). For LST acquisition, thermal infrared (TIR)-based remote sensing has attracted increasing attention due to the relatively fine spatial resolution, high accuracy and data availability across the globe (Li et al. 2013). Over the past decades, a wealth of LST products based on TIR algorithms have been developed and used, such as Moderate Resolution Imaging Spectroradiometer- (MODIS-), Spinning Enhanced Visible and Infrared Imager- (SEVIRI-), Advanced Very High-resolution Radiometer- (AVHRR-) and Visible Infrared Imaging Radiometer- (VIIRS-) LST. However, cloud contamination often leads to large gaps in TIR LST. For example, the MODIS LST covers less than 40% of the globe, which severely hinders the potential applications of the data (Duan et al. 2017). Therefore, filling LST gaps caused by cloud contamination is an urgent priority in relevant studies.

A series of approaches have been developed to fill the LST gaps and generate all-weather LST products (Shen et al. 2015; Wu et al. 2021). The early mainstream methods typically utilized the spatio-temporal information of TIR LST itself, considering LST temporal variation laws, such as the diurnal temperature cycle (DTC) (Xu and Shen 2013) and annual temperature cycle (ATC) (Xia et al. 2021), or spatial neighboring laws, such as kriging (Ke et al. 2013) and spline function (Neteler 2010), or both of them (Kilibarda et al. 2014; Li et al. 2018; Weiss et al. 2014). However, the accuracy of these methods is significantly affected by the accessibility of neighboring clear pixels in space and time, with an unsatisfactory performance found in large-scale or long time-series data missing cases (Fu and Weng 2016). Furthermore, this type of method cannot reflect the cloud effects, thus leading to a hypothetical clear-sky LST (Hong et al. 2021).

To obtain the real cloudy-sky LST, methods considering the difference between LST in clear and cloudy conditions were proposed. These methods normally take proxy data for clouds from other sources as the additional information, such as meteorological data, passive microwave (PMW)-based data, and land surface model (LSM)-simulated or assimilation data. By calculating and adding the cloud effect estimated from meteorological data, SEB-based methods have been proposed to correct the hypothetical clear-sky LST (Jia et al. 2021; Jin 2000; Zeng et al. 2018). Nevertheless, some meteorological data, such as ground-based air temperature and radiation data, are difficult to obtain, thus impeding the applications of SEB-based methods in poorly gauged regions (Lu et al. 2011; Yu et al. 2014b). Due to the ability to penetrate clouds, PMW LST is another commonly used data source for mapping real and gapless LST (Duan et al. 2017; Wu et al. 2022; Xu and Cheng 2021). However, the retrieval of high-quality PMW LST remains challenging due to the surface penetration, swath gaps and the relatively coarse spatial resolution, which can degrade the quality of the reconstructed LST (Zhang et al. 2020). Forced by all-weather meteorological data, LSMs can achieve the spatio-temporally continuous simulation of land surface parameters, such as LST, soil moisture (SM), and surface energy fluxes, and are regarded as a fundamental methodology in Earth system science (ESS) (Fisher and Koven 2020). In addition, LSMs are rooted in scientific theory based on physical mechanisms and parameterization schemes, thereby providing more realistic LST estimates with physical meanings. Some studies have incorporated LSM-simulated data as the complementary information to fill the gaps of TIR LST, which is an approach that have received much attention in recent years (Long et al. 2020; Ma et al. 2022; Zhang et al. 2021). However, LSM simulation data are usually of low accuracy due to the simplistic and static model representations of land-surface heterogeneities, which can introduce uncertainties in the reconstruction process (Trigo et al. 2015; Wang et al. 2014).

Motivated by the remarkable data mining and non-linear representation capabilities of ML architectures, many achievements have been made in LST reconstruction studies. The reconstruction is usually implemented by establishing the non-linear function between TIR LST and gapless LST (e.g., PMW LST) or its spatio-temporal descriptors, such as the normalized difference vegetation index (NDVI), digital elevation model (DEM), day of year (DOY), and albedo in clear conditions, and applying the relationship to obtain the cloudy LST (Buo et al. 2021; Tan et al. 2021b; Wu et al. 2022; Wu et al. 2019; Xiao et al. 2023). Thanks to the emergence and introduction of proxy data for clouds, ML methods have advanced the ability to reconstruct the actual cloudy-sky LST (Cho et al. 2022; Fu et al. 2019; Shwetha and Kumar 2016; Zhao and Duan 2020). Nevertheless, the mainstream methods incorporate independent proxy data (e.g., solar radiation data or LSM simulated LST) which are only correlated with cloudy LST, but provide no prior assumptions or physical understanding of the processes. As is well known, ML methods are empirically based and rely heavily on massive training data, making it difficult to transfer the model to other data-sparse regions (Lin et al. 2023). Furthermore, "correlation does not imply causation" (Aldrich 1995; Altman and Krzywinski 2015). The LST retrieval mechanisms and the process knowledge (such as conservation of energy) behind LST reasoning remain unclear, which may result in spurious predictions and extrapolation problems (Read et al. 2019). Therefore, it is necessary to incorporate causal data involved in the LST reasoning process as input, thereby coupling such mechanisms into the ML model to enhance its interpretability and transferability.

Coupling physics knowledge into ML is one of the current research hotspots, and has been successfully applied in estimating various land surface parameters (De Bézenac et al. 2019; Karniadakis et al. 2021; Karpatne et al. 2017; Koppa et al. 2022; Shang et al. 2023; Shen and Zhang 2023; Wang et al. 2021; Wang et al. 2023). Toward LST reconstruction, to the best of our knowledge, few studies have been conducted on coupling the mechanism and learning. Regarding the above issues, this study takes into consideration the rationalism and empiricism that inform a physics-constrained ML method for mapping gapless LST. Specifically, an advanced LSM, i.e., the Community Land Model (CLM), which uses meteorological forcing data to calculate the SEB was first used to produce model-based estimates. The complete process of CLM-LST modeling was deduced to identify the optimal combination of physical variables with causal relationships as inputs for the ML model. By incorporating key CLM forcing data (cause) and CLM simulation data (effect), the physical constraints based on the SEB underlying the data were integrated into the ML model. The generated LST product under all-weather conditions tends to be physically realistic while maintaining a high accuracy.

## 2. Study area and data

*2.1 Study area*

Fig. 1 shows the study area, which covers most of the middle and upper reaches of the Heihe River Basin (HRB-MU), and is located within 37.5°N–39.5°N and 99°E–101°E in arid Northwest China. Featuring glaciers, alpine meadows, grassland and forest in the upstream and dominated by irrigated crops and desert in the middle part of the HRB, the terrain of the HRB-MU is complex, with the elevation ranging from 1325 to 5040 m. The HRB-MU has a continental climate with mean annual precipitation of ~400 mm and mean annual temperature of ~275 K (Tan et al. 2021a). The in-situ measurements used in this study were derived from four weather stations in a well-known watershed observatory network, namely the Watershed Allied Telemetry Experimental Research (WATER) network (Li et al. 2009). The observatory network in the HRB-MU provides an ideal testbed for the validation of the proposed method in estimating gapless LSTs under various topographic conditions (Liu et al. 2018).

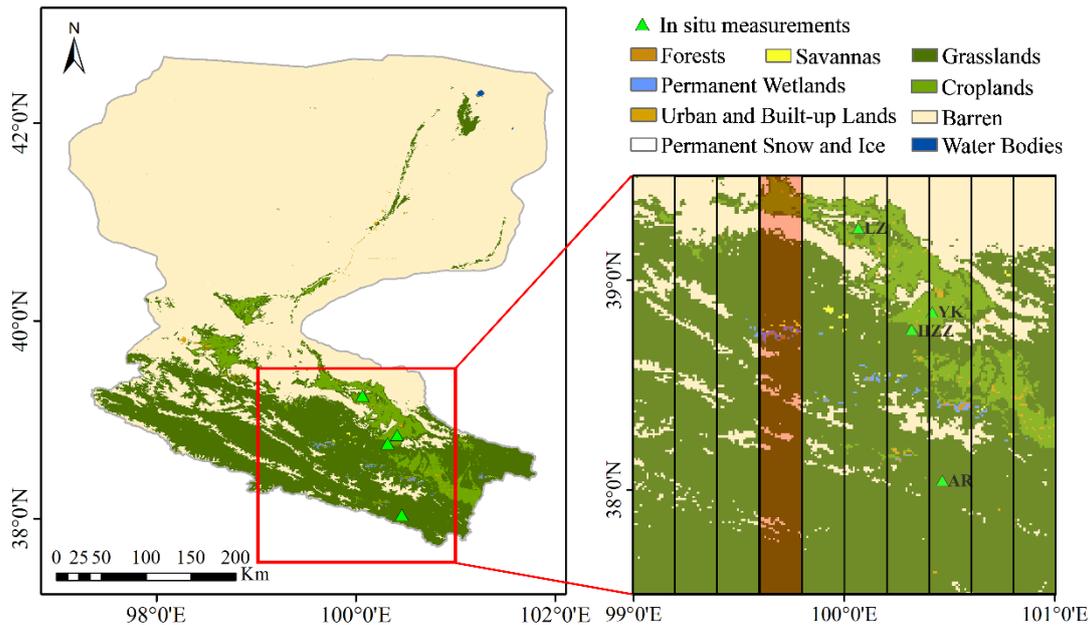

Fig. 1. Location of the Heihe River Basin (HRB) and the study area (HRB-MU). Land-cover types for 2010 are from the MODIS land-cover type product (MCD12Q1) with a 500m spatial resolution. The rectangle filled with light red represent the selected region for testing, while the other rectangles represent the regions for training in the space-based validation (Section 3.3).

*2.2. Data*

The data used in the study consisted of: 1) satellite data for model training and validation; 2) meteorological forcing data for model training, validation and CLM driving; 3) CLM simulations for model training and validation; and 4) in-situ LST observations for model validation. Before the model implementation, the satellite data, meteorological forcing data and CLM simulation data were all resampled to a 0.01° × 0.01° spatial resolution using a nearest neighbor interpolation and re-projected to the WGS84 coordinate system. Table 1 summarizes the basic information about the multi-source data used in this study.

**Table 1**

Summary of the satellite data, meteorological forcing data, model simulation data and in-situ measurements used in this study.[a]

| Short name | Variables | Spatial resolution | Temporal resolution | Usage |
|---|---|---|---|---|
| MYD11A1 | LST | 1 km | daily | Model label |
| MOD13A2 | NDVI | 1 km | 16-day | Model input |
| GLASS02A06 | Albedo | 1 km | 8-day | Model input |
| STAR NDSI | NDSI | 500 m | daily | Model input |
| SRTM DEM | DEM, latitude | 90 m | - | Model input |
| CLDAS forcing | TMP, RHU, PRS, SRA, WIN, PRE | 0.0625° | hourly | CLM driving Model input, |
| CLM simulated | CLM-LST, CLM-SM | 0.05° | hourly | Model input |
| WATER | ULW, DLW | point | 30-min | Model validation |

[a] The definitions of the variables can be found in the text and the download links are attached in the Acknowledgements.

*2.2.1 Satellite data*

The satellite data used in this study are listed in Table 1. The 1-km MODIS daily LST/emissivity product (MYD11A1) in Collection 6 was selected as the model label data. The MYD11A1 v6 product was retrieved from Aqua MODIS using the generalized split-window algorithm (Wan and Dozier 1996), with an approximate overpass time of 1:30 p.m. (local solar time) in ascending orbit and 1:30 a.m. in descending orbit. Some studies have indicated that the MODIS LST product has an accuracy of within 1 K in homogeneous areas (Wan 2014). The quality control (QC) flags with "LST produced, good quality (QC = 0)" and "average LST error <= 1 K" were used to identify the clear-sky MODIS LST data.

Four surface variables that are highly correlated with surface thermal properties were chosen as the LST predictors. The 16-day 1-km NDVI data were retrieved from the MODIS vegetation index product (MOD13A2). The Global LAnd Surface Satellite (GLASS) (Liang et al. 2021) black-sky surface albedo product with visible (B_VIS) and near-infrared (B_NIR) spectral ranges was obtained from GLASS02A06. A normalized difference snow index (NDSI) product produced by a Spatio-Temporal Adaptive fusion method with error correction (STAR-NDSI) (Jing et al. 2022) was adopted in this study.

The DEM and latitude (LAT) data were used to account for the LST terrain effect and temperature gradients (Minder et al. 2010; Zhao et al. 2019). We acquired the Shuttle Radar Topography Mission DEM (SRTM DEM) data with a 90-m spatial resolution from the United States Geological Survey (USGS). For the temporal feature, the DOY was selected.

The temporal resolutions of the GLASS product and MODIS vegetation index product vary from 8 days to 16 days. To obtain these products at a daily time scale, different interpolation methods were employed. The nearest neighbor interpolation approach was used for albedo due to its relative invariance within 8 days. As for the NDVI, the linear interpolation approach was applied, as it can represent the true trend of this variable. After the preprocessing, all the auxiliary variables were gapless with a 1-km and daily spatio-temporal resolution.

*2.2.2 Meteorological forcing data and surface data*

The China Meteorological Administration Land Data Assimilation System version 2.0 (CLDAS v2.0) (Shi et al. 2011) was used as the forcing data for driving the Community Land Model Version 5.0 (CLM5.0) and the input data for the ML model. The CLDAS v2.0 product was developed and maintained by the China Meteorological Administration (CMA), covering East Asia (0–65°N, 60–160°E) and spanning from 2008 to 2018. It was produced by merging reanalysis data, satellite-based products, and in-situ measurements from more than 2400 national automatic stations of the CMA, with a spatial resolution of $0.0625° \times 0.0625°$ and a temporal resolution of 1 h (Liu et al. 2019). The dataset consists of six meteorological forcing variables, i.e., 2-m air temperature (TMP), 2-m specific humidity (RHU), 10-m wind speed (WIN), precipitation rate (PRE), air pressure (PRS), and downward shortwave radiation (SRA). Compared with the Global Land Data Assimilation System (GLDAS) forcing data, the CLDAS product has been proven to be more accurate in China due to the integration of more site observations (Yang et al. 2017). The soil properties data, i.e., percent sand, percent clay and organic matter density with a 30 arc-second resolution, were derived from the China Dataset of Soil Properties for Land Surface Modeling (Shangguan et al. 2013). The static land-cover data, including percent crop/lake/wetland/glacier/urban/natural vegetation and different percent plant functional types (PFTs) were collected from the MODIS land-cover type product (MCD12Q1) for 2010. The other surface data, such as elevation, slope, and the monthly leaf and stem area index, were obtained from the CLM surface data pool (Oleson et al. 2010).

The above-mentioned high-quality forcing and surface data were introduced to replace the default datasets in the CLM5.0 data pool, to improve the model performance. After the CLM5.0 build and spin-up, a four-year numerical simulation was conducted to produce model-based LST and SM estimates with a 0.05° and 1-h spatio-temporal resolution. To match the satellite data, the CLDAS forcing data and CLM simulation data for ML model training were temporally interpolated to the value at the MYD11A1 view time by a cubic spline interpolation method.

*2.2.3 In-situ data*

To access the accuracy of the reconstructed LST, four in-situ longwave radiation measurements with different land-cover types from 2008 to 2011 were collected from WATER, who provided continuous and high-quality in-situ measurements for LST assessment studies (Duan et al. 2017; Wu et al. 2022; Yu et al. 2014a). The A'rou freeze station (AR), Huazhaizi desert station (HZZ), Yingke oasis station (YK), and Linze grassland station (LZ) are equipped with Kipp & Zonen (CNR1/CNR4) or CAMPBELL (CG3) net radiometers for measuring upwelling (ULW) and downwelling longwave radiation (DLW) every 30 min. Fig. 1 shows the locations of the stations, and Table 2 lists the basic information about the four stations.

The in-situ LSTs were retrieved using Stefan-Boltzmann law as follows (Wang and Liang 2009):

$$T_s = \left[\frac{F^\uparrow - (1-\varepsilon_b)F_\downarrow}{\varepsilon_b \cdot \sigma}\right]^{1/4} \tag{1}$$

where $T_s$ is the surface skin temperature or LST; $F^\uparrow$ and $F_\downarrow$ denote the surface upwelling and atmospheric downwelling longwave radiation, respectively; $\sigma$ is the Stefan-Boltzmann constant ($5.67 \times 10^{-8}$ W m$^{-2}$ K$^{-4}$); and $\varepsilon_b$ is broadband emissivity (BBE), which was acquired from a GLASS BBE product representing the emissivity value at 8–13.5 μm (Cheng et al. 2015). Finally, the "3σ (standard deviations) edit rule" method (Pearson 2002) was utilized to remove outliers due to cloud contamination.

**Table 2**

The basic information about the four weather stations used in this study.

| Site | Location | Elevation (m) | Land-cover | Period |
| --- | --- | --- | --- | --- |
| AR | 100.4647°E, 38.0444°N | 3033 | Alpine meadow | 2008/01–2011/12 |
| HZZ | 100.3201°E, 38.7659°N | 1731 | Desert steppe | 2008/01–2011/12 |
| YK | 100.4167°E, 38.85°N | 1519 | Cropland | 2008/01–2011/12 |
| LZ | 100.0667°E, 39.25°N | 1394 | Grassland | 2008/01–2008/10 |

## 3. Methodology

We propose a physics-constrained ML model for mapping gapless LST. The hybrid model takes ML as the main architecture, under which the underlying physical laws (e.g., the SEB) between the key CLM forcing data (SRA, TMP, RHU, PRS) and CLM simulation data (CLM-LST, CLM-SM) in the CLM-LST modeling are incorporated and constrain the ML model. The LST estimations of the hybrid model were compared with those of a pure physical method, pure ML methods, and other advanced ML methods. The well-trained model was applied for the HRB-MU from 2008 to 2011. The generated LST data were validated against in-situ LST and compared with other all-weather LST data. The details of the proposed method are provided below.

*3.1 Pure ML model*

ML has gained increasing popularity in gapless LST estimation (Li et al. 2021; Zhang et al. 2020; Zhao and Duan 2020). To date, the ML has usually been implemented by establishing the relationship between the clear-sky LST and auxiliary variables. The established relationship was then subsequently applied to the gapless auxiliary variables for mapping all-weather LST. The early studies normally used gapless data (e.g., remote sensing data) as the explanatory variables. The empirical model can be constructed as follows:

$$LST = \mathbf{ML}(DEM,\ LAT,\ NDVI,\ NDSI,\ B\_VIS,\ B\_NIR,\ DOY) \quad (2)$$

where $\mathbf{ML}$ is the pure ML model without physical constraints; $LST$ is the MODIS LST with the highest quality; $DEM, LAT, NDVI, NDSI, B\_VIS$, and $B\_NIR$ are the remote sensing data, which represent the surface thermal properties and terrain effect of LST; and $DOY$ represents the temporal characterization.

In this study, the light gradient-boosting machine (LGBM) model was used as the learner. Fig. 2 shows a schematic diagram of the LGBM model. As the successor of the extreme gradient boosting (XGBoost) model, the LGBM model is a newly developed ML algorithm based on the gradient-boosting decision tree (GBDT) (Ke et al. 2017). Compared with conventional deep learning methods, such as a deep belief network (DBN) and generalized regression neural network (GRNN), the LGBM model is advantageous in dealing with massive data and countering overfitting problems. A second-order approximation is used to minimize the objective function in the LGBM model, which is formulated as follows:

$$L^{(t)} = \sum_{i=1}^{n} \left[ l(y_i, \hat{y}^{(t-1)}) + \partial_{\hat{y}^{(t-1)}} l(y_i, \hat{y}^{(t-1)}) f_t(x_i) + \frac{1}{2} \partial_{\hat{y}}^2(t-1) l(y_i, \hat{y}^{(t-1)}) f_t^2(x_i) \right] + \Omega(f_t) \quad (3)$$

where $L^{(t)}$ is the objective function of the $t-th$ iteration solution; $x_i$ is the sample; $n$ denotes the number of samples; $l$ denotes the loss function used to measure the difference between the actual value $y_i$ and the predicted value $\hat{y}$, which is the root-mean-square error (RMSE) in this case; $\partial$ and $\partial^2$ denote the first- and second-order gradients of $l$, respectively; $f_t$ is an independent regression tree and $f_t(x)$ is the corresponding increment; and $\Omega(f)$ denotes the regularization term, which is defined as follows:

$$\Omega(f) = \gamma T + \frac{1}{2}\lambda \|w\|^2 \qquad (4)$$

where $T$ denotes the number of leaf nodes and $w$ denotes the leaf weights, which are adopted from a leaf-wise tree growth strategy.

The effectiveness of the LGBM model mainly stems from Gradient-based One-Side Sampling (GOSS) and Exclusive Feature Bundling (EFB) (Fig. 2). GOSS is a novel sampling technique that keeps a good balance between reducing the number of data instances and maintaining a high accuracy. Firstly, it sorts the data instances based on their gradient absolute values and selects the top m % instances. Secondly, it randomly samples n % instances from the remaining data. Finally, the sample instances with small gradients are amplified by a $\frac{1-m\%}{n\%}$ weight to ensure that more emphasis is placed on the under-trained instances without changing the original data distribution. EFB was designed to safely filter the features (such as NDVI, NDSI, B_VIS, B_NIR, etc.) based on the sparsity of high-dimensional data, which is founded on a histogram-based algorithm. It first clusters the data to find those sparse but frequently co-occurring features, which are usually mutually exclusive. It then bundles these exclusive features into a single feature to avoid unnecessary computation for zero feature values. This bundling process can reduce the dimension of the samples, thereby reducing memory usage, preventing overfitting and retaining accuracy.

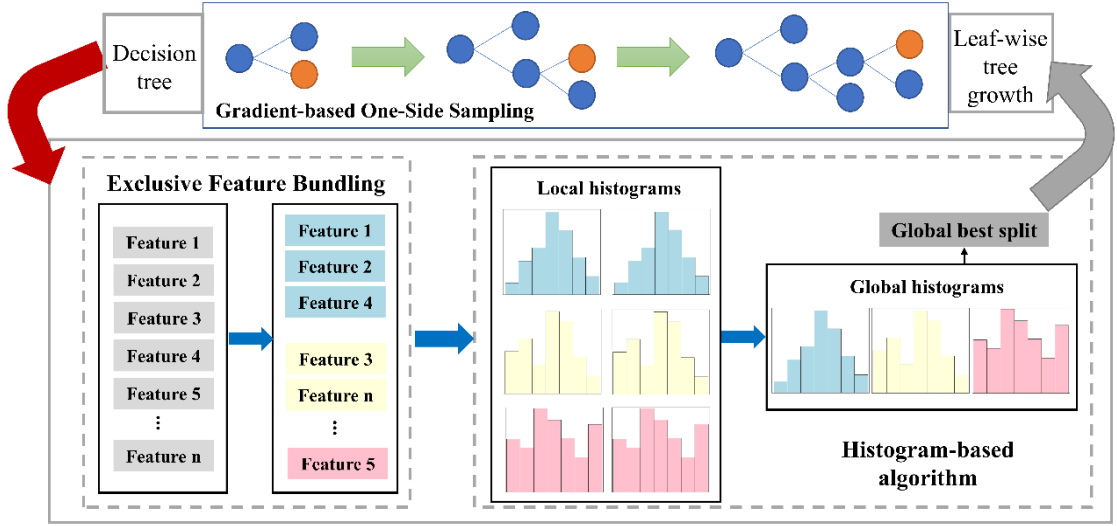

Fig. 2. Schematic of the light gradient-boosting machine model.

*3.2 Physics-constrained ML model*

*3.2.1. Overall framework*

Despite filling the gaps of the MODIS LST, the obtained cloudy LST based on pure ML (Eq. (2)) is far away from realistic due to the unavailability of data that can distinguish between clear-sky and cloudy-sky conditions. Furthermore, pure ML methods can only learn the relationship between discrete LST samples and the explanatory data. The causal relationship and the underling physical process between LST retrieval input and output remain unclear. As a result, the obtained LST lacks physical interpretability, which is usually ignored in LST reconstruction studies using ML methods. In our previous work, three constraint methods were proposed to integrate the physical knowledge into the ML model, i.e., input variable constraints, objective function constraints, and model structure constraints (Shen and Zhang 2023). In this case, the input variable physical constraints are employed. Specifically, the key CLM forcing data, CLM simulation data and auxiliary data are jointly input into the LGBM, which forms an LST retrieval closed-loop. The relationship is improved as:

$$LST = \mathbf{\textit{PC-LGBM}}(\mathbf{\textit{CF}}, \mathbf{\textit{CS}}, DEM, LAT, NDVI, NDSI, B\_VIS, B\_NIR, DOY) \quad (5)$$

where $\mathbf{\textit{CF}} = \{SRA, TMP, RHU, PRS\}$ denotes the four sets of CLM forcing data that are physically dominant in CLM-LST simulation; and $\mathbf{\textit{CS}} = \{CLM\text{-}LST, CLM\text{-}SM\}$ represents the two sets of CLM simulation data that correspond to the CLM forcing data. Both $\mathbf{\textit{CF}}$ and $\mathbf{\textit{CS}}$ can characterize the real thermal state under clouds. The downward shortwave radiation *SRA* is only used in the daytime model. ***PC-LGBM*** is the physics-constrained ML model obtained by imposing SEB constraints to the LGBM model (Fig. 3). The LST retrieval mechanisms and the involved process knowledge in CLM5.0 are deduced and given below.

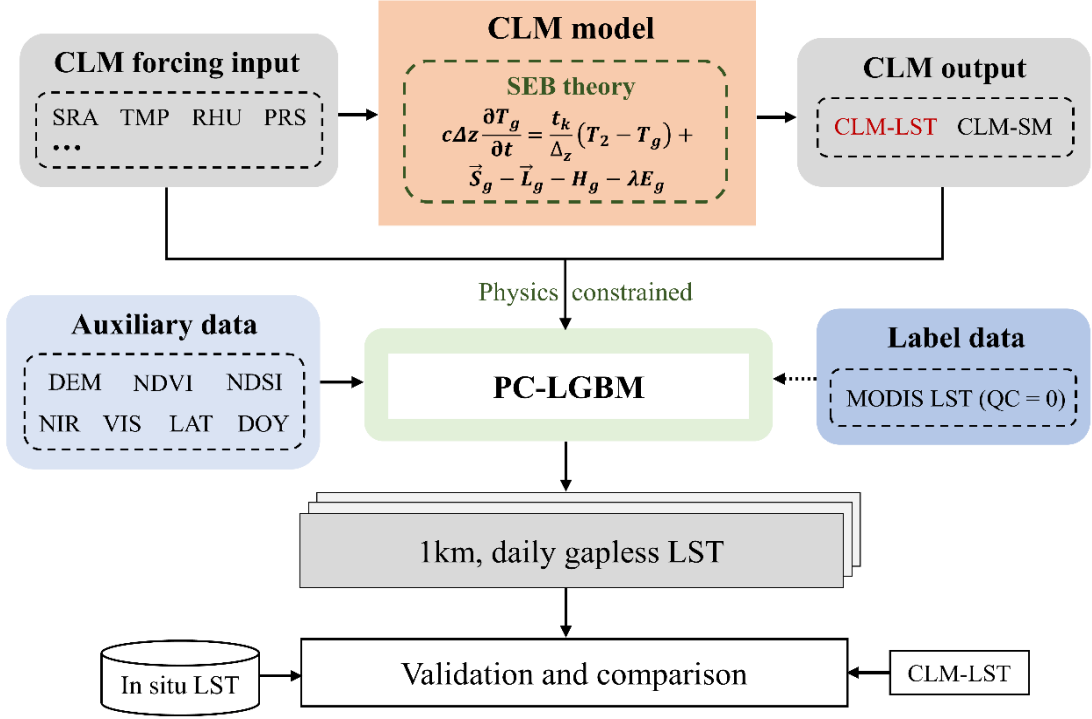

Fig. 3. Framework of the proposed physics-constrained ML method for mapping gapless LST. $\Delta z$ is the layer thickness between two layers; $T_g$ and $T_2$ are the soil temperatures at the first and second layer, respectively; $\vec{S}_g$ and $\vec{L}_g$ are the solar radiation and the net longwave radiation absorbed by the ground, respectively; and $H_g$ and $\lambda E_g$ are the sensible and latent heat fluxes, respectively.

*3.2.2. The CLM5.0 model and its configuration*

In this study, the Community Land Model Version 5.0 (CLM5.0) was utilized to produce the model-based LST and SM estimates. CLM5.0 is the land component coupling in the Community Earth System Model 2 (CESM2, https://www.cesm.ucar.edu/models/cesm2), which is the latest version released by the National Center for Atmospheric Research (NCAR) in 2019 (Lawrence et al. 2019). Compared with CLM4.5, CLM5.0 is updated with several parameterizations in soil and plant hydrology, snow density, river modeling, carbon and nitrogen cycling and crop modeling (Lawrence et al. 2019). Previous studies have confirmed that CLM5.0 performs better than CLM 4.5 in soil temperature simulation (Deng et al. 2020; Luo et al. 2020). A 7-year (2001–2007) spin-up simulation at the regional scale over the HRB-MU was implemented to reach an initialization equilibrium of the thermal regime for CLM5.0. Subsequently, offline numerical simulations with a 0.05° × 0.05° grid resolution and a 1-h time interval during 2008–2011 were conducted in the prescribed satellite phenology mode.

*3.2.3. Mechanistic reasoning of the physical constraints coupled in the ML model*

As mentioned, the input variable constraints are applied to incorporate physical knowledge into the LGBM model. Thus, it is important to explore which forcing data are used to carry out the mechanism simulation of the dynamics. Here, we provide a detailed reasoning process for the retrieval of LST in the CLM, to identify the meteorological forcing data that are physically related to CLM-LST. The mathematical derivation is given below.

In CLM 5.0, LST is computed as follows:

$$T_s = \left(\frac{L\uparrow}{\sigma}\right)^{1/4} \tag{6}$$

where $T_s$ is the skin temperature (LST); and $L\uparrow$ is the surface upward longwave radiation, which can be estimated as:

$$L\uparrow = \begin{cases} (1-\varepsilon_g)L_{atm}\downarrow + \varepsilon_g \sigma(T_g)^4, & L+S < 0.05 \\ (1-\varepsilon_g)(1-\varepsilon_v)^2 L_{atm}\downarrow + \varepsilon_v[1+(1-\varepsilon_g)(1-\varepsilon_v)]\sigma(T_v)^4 + \varepsilon_g(1-\varepsilon_v)\sigma(T_g)^4, & \text{otherwise} \end{cases} \tag{7}$$

where $L$ and $S$ denote the exposed leaf area index and stem area index, respectively; $L+S < 0.05$ represents the non-vegetated surfaces, and otherwise vegetated surfaces; $T_g$ and $T_v$ are the ground (soil, snow and surface water) and vegetation temperatures, respectively; $\sigma$ is the Stefan-Boltzmann constant; $\varepsilon_g$ and $\varepsilon_v$ are the ground and vegetation emissivity, respectively; and $L_{atm}\downarrow$ is the downward atmospheric longwave radiation, which is calculated based on Idso (1981):

$$L_{atm}\downarrow = \left[0.7 + 5.95 \times 10^{-5} \times 0.01 \times e_{atm} \times exp(\frac{1500}{T_{atm}})\right] \sigma T_{atm}^4 \tag{8}$$

where $e_{atm} = \frac{P_{atm}q_{atm}}{0.622+0.378q_{atm}}$ is the atmospheric vapor pressure; and $P_{atm}$, $q_{atm}$, and $T_{atm}$ are the air pressure (PRS), the air specific humidity (RHU), and the air temperature (TMP), which were obtained from the CLDAS meteorological forcing dataset.

The numerical solutions for $T_g$ and $T_v$ are based on the SEB equation. Taking $T_g$ as an example, the SEB is defined as follows:

$$G = \vec{S}_g - \vec{L}_g - H_g - \lambda E_g \tag{9}$$

where $G$ is the ground heat flux, which is used for the soil temperature calculation; $\vec{S}_g$ and $\vec{L}_g$ are the solar radiation and the net longwave radiation absorbed by the ground, respectively; $H_g$ and $\lambda E_g$ are the sensible and latent heat fluxes, respectively, where $E_g$ is the water vapor flux, and $\lambda$ is a multiplier for converting the water vapor flux to an energy flux. The surface net radiation ($\vec{S}_g - \vec{L}_g$) is given as:

$$R_{n,g} = (1-\alpha)S\downarrow + \varepsilon_g L_{atm}\downarrow - \varepsilon_g \sigma (T_g)^4 \tag{10}$$

where $R_{n,g}$ is the surface net radiation; $S\downarrow$ is the incident solar flux (W m$^{-2}$), which is obtained from the solar shortwave radiation (SRA); and $\alpha$ is the ground albedo, which is associated with the land-cover type and soil color.

The calculations of $H_g$ and $E_g$ are based on Monin-Obukhov similarity theory (Monin and Obukhov 1954) using the ground temperature from the previous time step, in conjunction with the atmospheric potential temperature, specific humidity, and thermodynamic and aerodynamic resistances. The formulas for $H_g$ and $E_g$ can be found in CLM5.0 technical note (https://escomp.github.io/ctsm-docs/versions/release-clm5.0/html/tech_note/). $H_g$ and $E_g$ are then used as the surface forcing for the solution of the ground temperature at the current time step. The law of heat conduction in one-dimensional form is:

$$c\frac{\partial T}{\partial t} = \frac{\partial}{\partial z}\left[t_k \frac{\partial T}{\partial z}\right] \tag{11}$$

where $c$ denotes the volumetric heat capacity of snow/soil (J m$^{-3}$ K$^{-1}$); $t$ denotes the time, which was 3600 s in this study; $t_k$ denotes the thermal conductivity (W m$^{-1}$ K$^{-1}$), depending on the soil property; and $z$ denotes the vertical direction depth (m). Combining Eq. (9) and Eq. (11) yields the energy balance equation for LST calculation:

$$c\Delta z \frac{\partial T_g}{\partial t} = \frac{t_k}{\Delta_z}(T_2 - T_g) + \vec{S}_g - \vec{L}_g - H_g - \lambda E_g \tag{12}$$

where $\Delta z$ denotes the layer thickness (m) between the two layers; $T_g$ and $T_2$ are the soil temperature (K) at the first and second layer, respectively. In CLM5.0, there are 25 layers in total, which are thinker from the top layer to the bottom layer. This equation is solved numerically using the Crank-Nicolson method to calculate the ground temperature, with the boundary conditions of $G$ as the ground heat flux into the top ground surface from the overlying atmosphere.

According to Eq. (12), LST is determined by subtracting the outgoing energy from the incident energy, which represents the amount of energy that is absorbed by the surface (Jin 2000). To represent the incident energy, the SRA is used (Eq. (10)). In CLM5.0, the downward atmospheric longwave radiation $L_{atm}\downarrow$ also exerts a crucial role on LST (Eq. (7) and Eq. (10)). The $L_{atm}\downarrow$ is parameterized using TMP, PRS, and RHU (Eq. (8)). Furthermore, TMP and RHU are also required in the sensible and latent heat flux parameterizations. In this respect, the CLM forcing data (specifically, SRA, TMP, PRS and RHU) with a strong constraint relationship on CLM-LST are selected as the input variables for the PC-LGBM model, making it feasible in the identification of causation from the correlation between the model forcing data and simulation data.

*3.3 Evaluation strategies*

LST varies significantly on diurnal and intra-annual scales (Göttsche and Olesen 2001; Zhan et al. 2014). In this respect, we divided all the samples from 2008–2011 into eight data subsets based on spring (March–May), summer (June–August), autumn (September–November), and winter (December–February), as well as daytime and nighttime. Therefore, eight models in total were trained and evaluated.

Sample-based validation, space-based validation and independent site validation were used for the model validation. For the sample-based validation, 10-fold cross-validation (CV) was utilized, i.e., all the samples were randomly divided into 10 folds, with nine folds were used for model training and one for model validation. This process was repeated 10 times and the average accuracy was calculated. The sample-based CV shows superiority in evaluating the model's precision and overfitting problems (Rodriguez et al. 2009). However, it is less effective in tackling extrapolation problems caused by the uneven sample distribution (Shen et al. 2022). Accordingly, space-based validation was applied to further test the model's generalization ability in the untrained sample space. Specifically, the study area was evenly divided into 10 sub-regions (Fig. 1). Samples from each sub-region were separately regarded as a test set, with other samples as the training set, to test the model's performance in regions without valid observations. The generated LST product was then further compared with independent in-situ LST and the CLM-LST. The Pearson correlation coefficient (R), the RMSE, the mean absolute error (MAE) and the overall bias (BIAS) were selected as the validation indicators.

**4 Results**

*4.1 Spatio-temporal patterns of the PC-LGBM estimated gapless LST*

One representative year (2010) was selected to enable the investigation into the spatio-temporal patterns of the generated gapless LST. Fig. 4 shows the spatial distribution of the CLM-simulated LST, MODIS LST, and estimated gapless LST in the different seasons (DOYs 16, 105, 196, and 287) in the daytime and nighttime of 2010. Due to the interference of clouds, missing data are prevalent in the MODIS LST. Despite possessing spatial completeness, the CLM-LST has relatively few spatial details. Comparatively, the PC-LGBM LST is spatially complete and retains a high consistency with the MODIS LST in spatial details as well as LST magnitude. In addition, there are no block effects or artifacts in the estimated LST image, indicating that the PC-LGBM model can successfully address the scale inconsistencies between the input data and MODIS data. As shown in Fig. 4, the LST is highly variable in space and time, especially during the summer daytime, whereas relatively gentle variation is found during the nighttime. Relatively high LST values are observed in the low-altitude northeast region of the HRB-MU, where bare land is dominant. Likewise, relatively low LST values are observed in the high-altitude southern HRB-MU, where grassland and forest are widespread. Overall, it is viable to use the PC-LGBM model to generate gapless LST with full spatial continuity and to accurately depict the spatio-temporal thermal dynamics of the land surface.

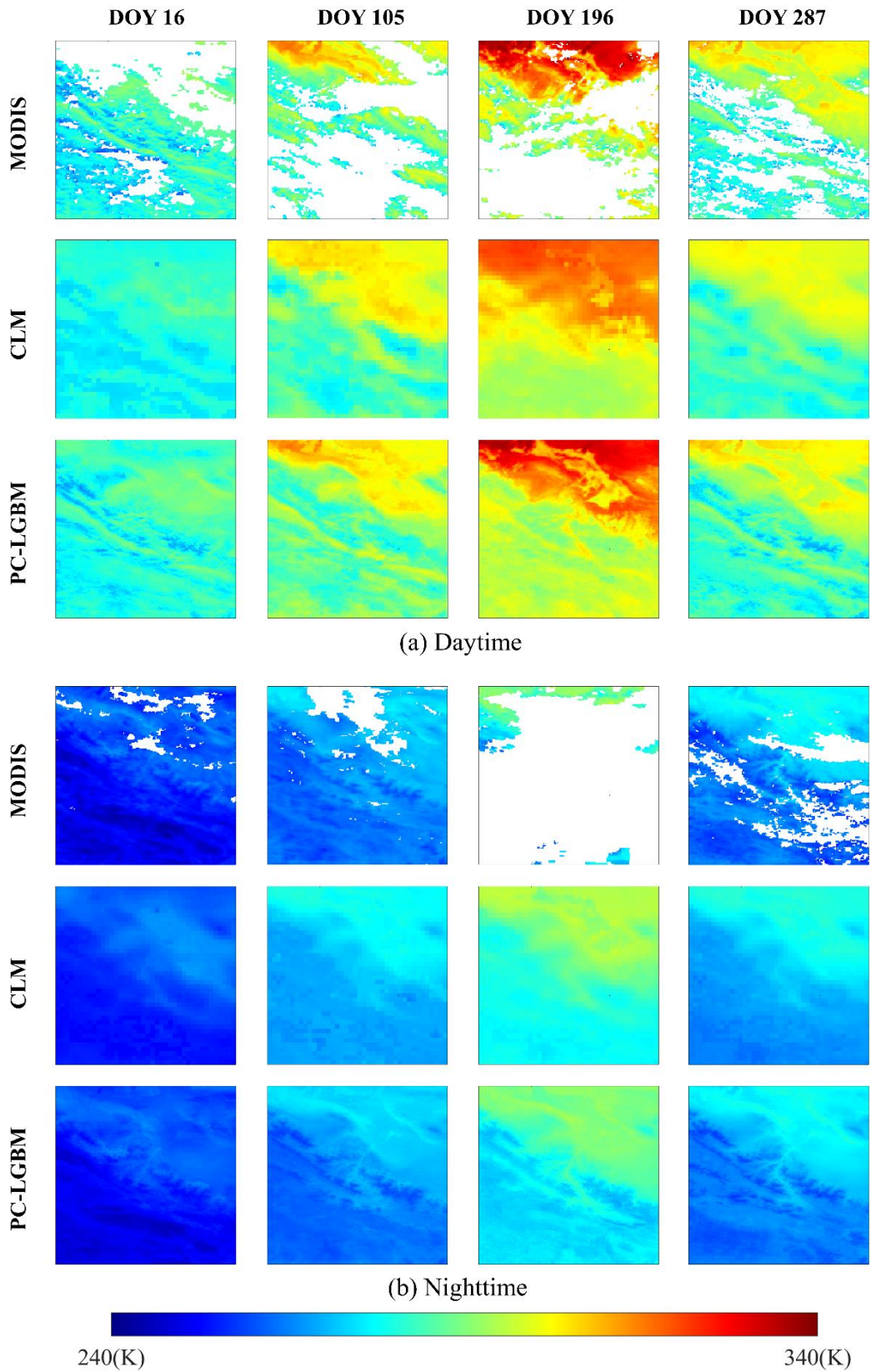

Fig. 4. Spatial patterns of the CLM-simulated LST, original MODIS LST, and estimated gapless LST for DOYs 16, 105, 196, and 287 in 2010 during (a) the daytime (around 13:30 local time) and (b) nighttime (around 01:30 local time).

Fig. 5 displays the temporal variation of the estimated all-weather LST during the daytime and nighttime in 2010 over the three sites. The in-situ LST, CLM-simulated LST, and the corresponding MODIS LST are also provided for a comparison. The black, blue, and red lines represent the in-situ LST, CLM-simulated LST, and estimated all-weather LST, respectively. The clear-sky MODIS LST is represented as a black hollow circle. The results show that the estimated all-weather LST can basically capture the seasonal and daily variations during both the daytime and nighttime, with an accuracy comparable to that of clear-sky MODIS LST, but with stronger time continuity. It is shown that the estimated daytime LST varies more sharply than the nighttime LST, with an RMSE of 2.92–3.79 K and 2.49–2.66 K during the daytime and nighttime, respectively. Benefiting from the incorporation of the physical law constraints, the proposed method extrapolates well and can capture some of the sudden rises (e.g., DOY 155 during the daytime) or drops (e.g., DOY 67 during the nighttime) in LST caused by extreme weather. Compared with the estimated LST, the CLM-LST exhibits an inferior performance, in terms of an RMSE of 3.28–6.22 K. Large warm biases can also be observed in CLM-LST during the nighttime, especially over bare land and sparsely vegetated regions (e.g., AR and HZZ). Previous studies have also reported this inherent warm nighttime bias in LSM simulated LST (Trigo et al. 2015; Wang et al. 2014). Through the powerful ability of the ML model in data mining, the estimated LST is to largely compensate for this overestimation.

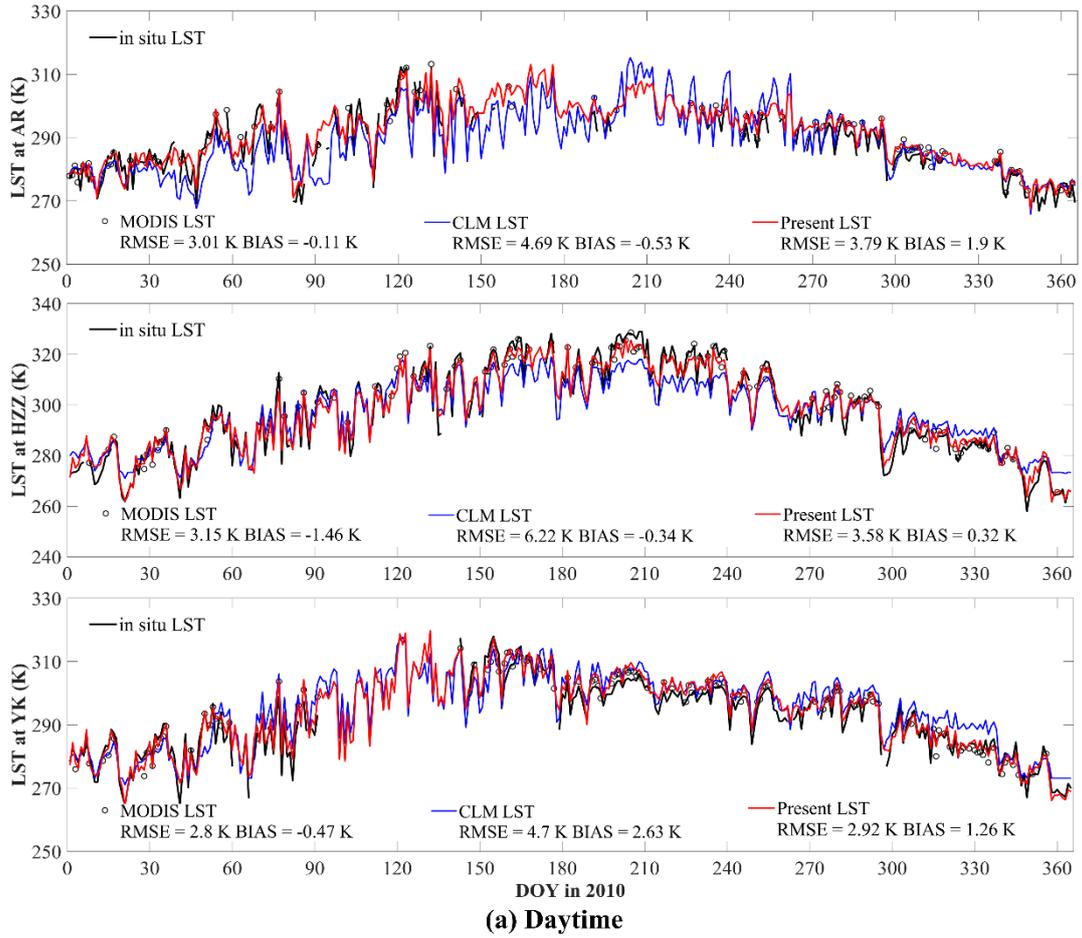

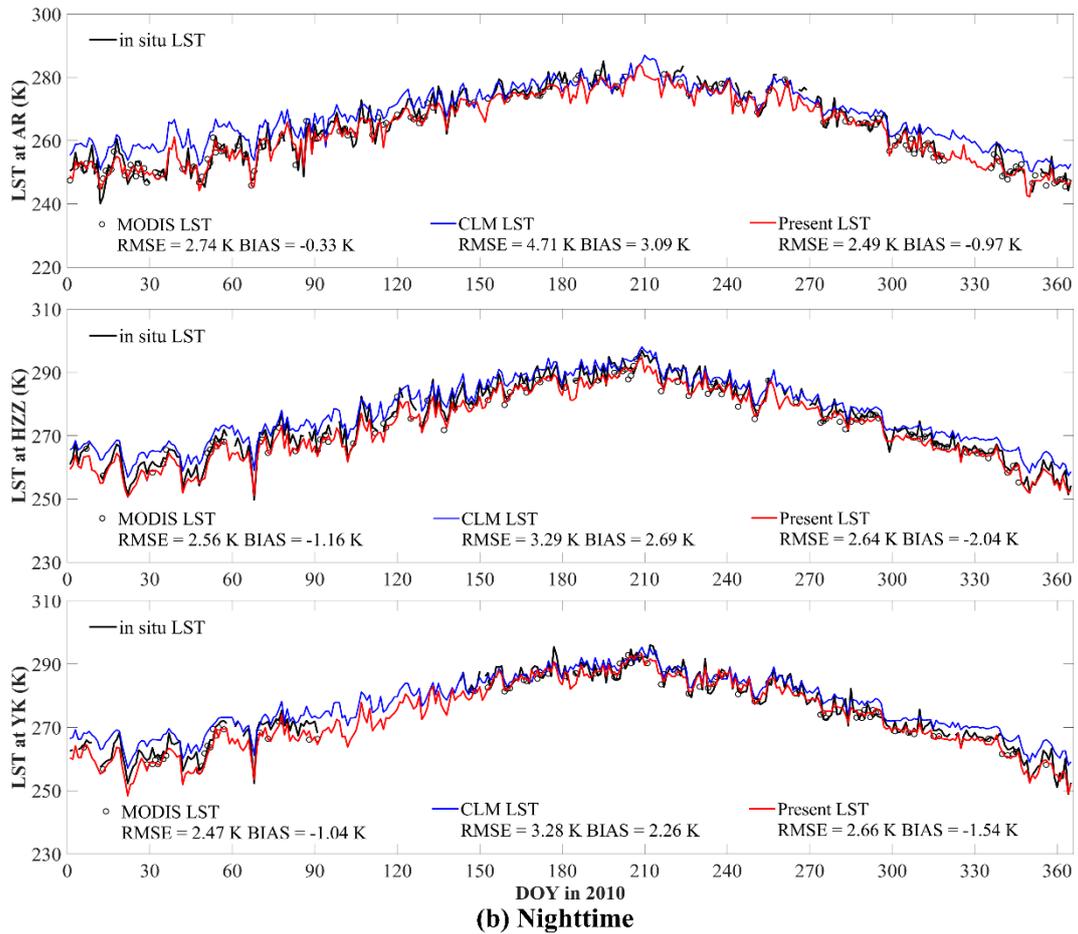

Fig. 5. Time series of the in-situ LST, clear-sky MODIS LST, CLM-simulated LST, and the estimated all-weather LST at the three sites in 2010 during (a) the daytime and (b) the nighttime. The first, second and third rows in each subgraph represent the AR, HZZ and YK sites, respectively. DOY denotes the day of year. The discontinuity of the black line is mainly due to the missing in-situ LST data.

*4.2 Quantitative evaluation results*

The performance of the PC-LGBM model was validated by simulation and real-data experiments using MODIS LST and in-situ LST measurements, respectively. Fig. 6 shows the sample-based and space-based validation results for the daytime and nighttime. As witnessed in Fig. 6, the sample size of the validation is adequate (>10 million), which demonstrates the reliability of the validation results. For the sample-based CV results, the averages of the R, RMSE, MAE, and BIAS are 0.99, 2.03 K, 1.49 K, and 0 K, respectively, for the daytime. Meanwhile, these values for the nighttime are 0.99, 1.23 K, 0.92 K, and 0 K, respectively. The results indicate that the PC-LGBM model shows a favorable performance in LST prediction. Moreover, the PC-LGBM model shows good results under the space-based validation, with an R of 0.98 (0.99), RMSE of 2.62 K (1.56 K), MAE of 1.94 K (1.16 K), and BIAS of −0.05 K (0.02 K) for the daytime (nighttime), which verifies the predictive ability of the model in regions without training samples. Compared with the model performance in the daytime, a higher accuracy can be observed in the nighttime, which could be associated with the high thermal heterogeneity and TIR directional anisotropy during daytime (Cao et al. 2019; Zhang et al. 2021).

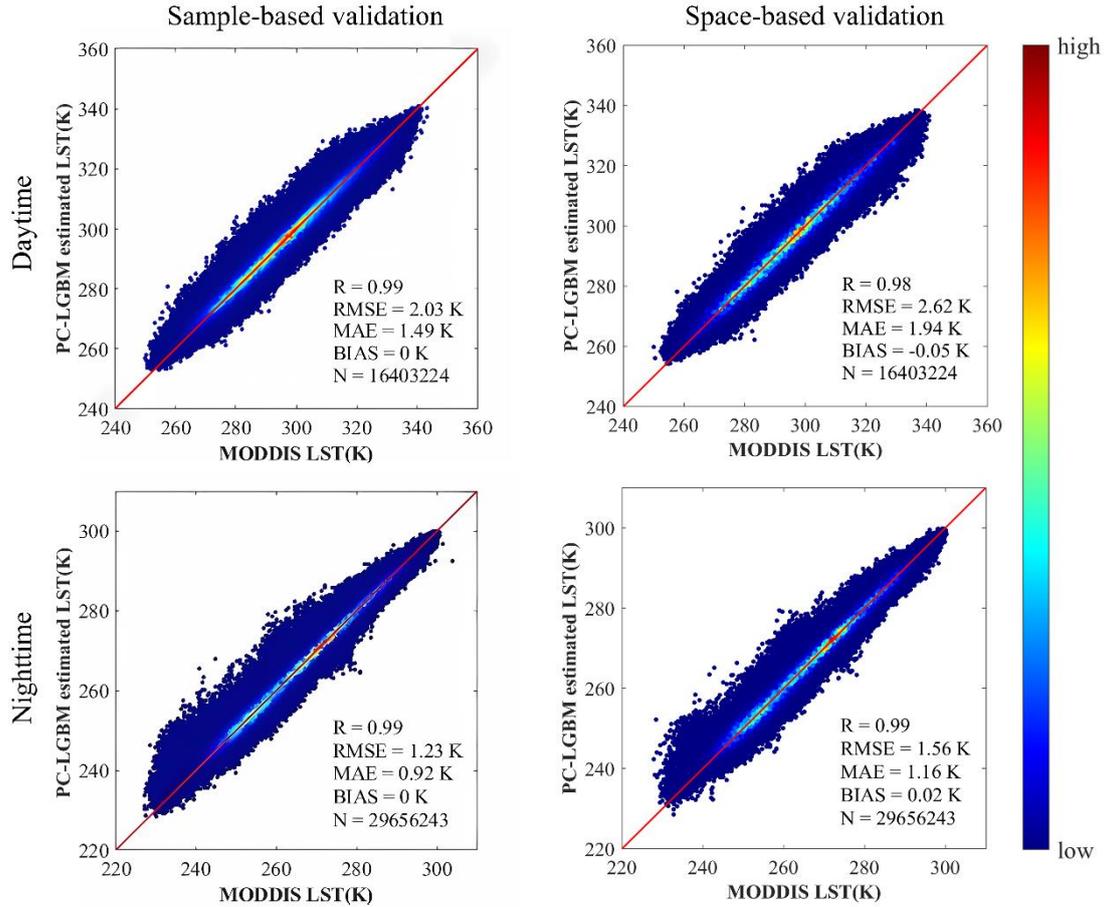

Fig. 6. Scatter plots of the sample-based CV results and space-based CV results for the daytime and nighttime. The red solid line denotes the 1:1 line. The color bar is the density of the samples. N represents the sample size.

In-situ LST measurements from four sites with different land-cover types were further used to assess the estimated gapless LST. A comparison of the estimated LST with in-situ LST measurements under all-weather conditions is shown in Fig. 7. As can be seen, the PC-LGBM estimated LST is in good agreement with the in-situ LST under clear-sky conditions, with an RMSE of 2.45–2.79 K, MAE of 2.05–2.21 K and R of 0.99. In addition, the accuracy of the cloudy-sky LST is generally comparable to that of the clear-sky LST, in terms of RMSE differences of 0.3–0.89 K, suggesting that the proposed approach can successfully recover cloudy LST from the established clear-sky model. In more detail, a better performance is observed at the YK and LZ sites, with an RMSE of 2.91 K and 3.12 K, respectively, whereas a relatively poor performance is observed at the HZZ and AR sites, with an RMSE of 3.23 K and 3.66 K, respectively. The inferior accuracy can be attributed to the errors of the MODIS LST retrievals arising from the uncertainty in emissivity estimation in specific land-covers (e.g., the HZZ site in a desert region, and the AR site in an alpine snowy region) (Duan et al. 2019; Li et al. 2019). In general, the estimated gapless LST resembles the in-situ LST reasonably well under both clear-sky and cloudy-sky conditions, demonstrating the reliability of the proposed PC-LGBM for estimating all-weather LST.

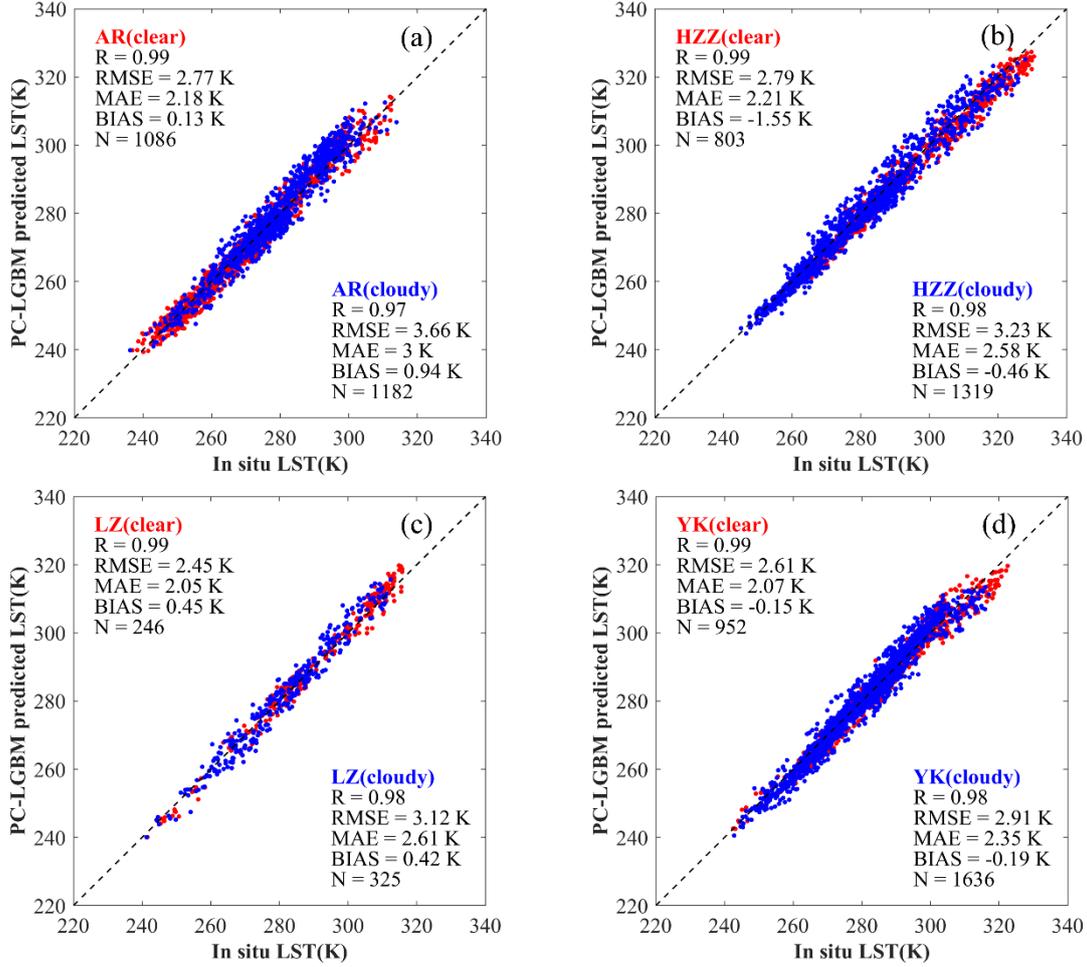

Fig. 7. Scatter plots of the estimated LST against in-situ LST measurements under clear-sky and cloudy-sky conditions during 2008 to 2011 at the four sites: (a) AR, (b) HZZ, (c) LZ, and (d) YK.

*4.3 Comparison with other methods and datasets*

*4.3.1 Comparison with pure ML methods*

In this study, the key CLM forcing data and CLM simulation data were jointly compiled for training an ML model, according to their strong causal associations. In order to better understand the contributions of physical constraints made to the LGBM model, the performance of ML models with different combinations of data was evaluated. Table 3 summarizes the model accuracy metrics for different combinations of data under the space-based validation. It should be clarified that RS, RS+CF, and RS+CS belong to pure ML methods, while RS+CF+CS belongs to a physics-constrained ML method (see Section 3.2). The results show that all the methods achieve reasonable accuracies, with the RMSE (MAE) lower than 3.88 K (2.94 K) and R greater than 0.97. In more detail, the worst accuracy is found when using only remote sensing data as input, with an RMSE of 3.88 K and 2.62 K for the daytime and nighttime, respectively. It is demonstrated that the daily remote sensing data interpolated from an 8-day or 16-day temporal resolution cannot sufficiently depict the high daily fluctuations in LST. When instantaneous key CLM forcing data or CLM simulation data are incorporated into the ML model, obvious improvements can be observed, compared with the basic result, followed by an RMSE decrease of 0.77–1.18 K, a MAE decrease of 0.6–0.94 K, and an R increase of 0.01–0.02. Compared with the pure ML methods, it is evident that the proposed PC-LGBM model (i.e., RS+CF+CS) performs the best among all the data combinations, which demonstrates the superiority of coupling physical knowledge with the ML model in LST estimation.

**Table 3**

Space-based model accuracy for different combinations of data, based on the LGBM model.

| Time | Input combinations | RMSE (K) | MAE (K) | R |
|---|---|---|---|---|
| Daytime | RS | 3.88 | 2.94 | 0.97 |
| | RS+CF | 2.7 | 2 | 0.98 |
| | RS+CS | 2.95 | 2.19 | 0.98 |
| | RS+CF+CS (PC-LGBM) | **2.62** | **1.94** | **0.98** |
| Nighttime | RS | 2.62 | 1.96 | 0.98 |
| | RS+CF | 1.65 | 1.22 | 0.99 |
| | RS+CS | 1.85 | 1.36 | 0.99 |
| | RS+CF+CS (PC-LGBM) | **1.56** | **1.16** | **0.99** |

The bold metrics denote the best performances among all the data combinations.

RS: remote sensing data; CF: key CLM forcing data, i.e., SRA, TMP, PRS and RHU; CS: CLM simulation data, i.e., CLM-LST and CLM-SM. The numbers of samples are 16403324 (29656243) for the daytime (nighttime), respectively.

To further evaluate the influence of the physical constraints on the estimated LST under all-weather conditions, scatter plots of the all-weather LST with different data inputs against in-situ LST are presented in Fig. 8. The CLM modeling is also added as a pure physical method for comparison. The accuracy of CLM-LST in cloudy-sky conditions (RMSE = 4.56 K, R = 0.97) is comparable with that under clear-sky conditions (RMSE = 4.63 K, R = 0.98), indicating that the model forcing and simulation data can effectively characterize the actual thermal information under clouds (Fig. 8a). This phenomenon also provides evidence in support of the reliability of applying the clear-sky ML model to recover LST in cloudy-sky conditions. Among all the estimated LSTs, the RS LST performs worst, with an RMSE of 3.57 K (6 K) and R of 0.94 (0.98) in clear-sky (cloudy-sky) conditions. In addition, it demonstrates a poor performance in reflecting the cooling effect caused by clouds, resulting in an overall overestimation (BIAS = 1.06 K) in LST in cloudy conditions (Fig. 8b). Once the model forcing data or simulation data are incorporated into the ML model, significant improvements in LST accuracy can be found, with an RMSE, MAE, BIAS, and R of 2.78–3.43 K, 2.19–2.73 K, −0.49–0 K, and 0.98–0.99, respectively (Fig. 8c−d). The best accuracy is found for RS+CF+CS LST, for which the RMSE, MAE, BIAS, and R are 2.7−3.23 K, 2.14−2.6 K, −0.37−0.07 K, and 0.98-0.99, respectively, which ranks first under both clear-sky and cloudy-sky conditions in all the scenarios (Fig. 8e). This again confirms that integrating key LSM forcing data and LSM simulation data simultaneously into the ML model actually makes sense in all-weather LST estimation. A reasonable illustration is that the PC-LGBM model is capable of implicitly learning the explicit process knowledge (e.g., the SEB) embedded in the CLM, constraining the learning process and improving the accuracy.

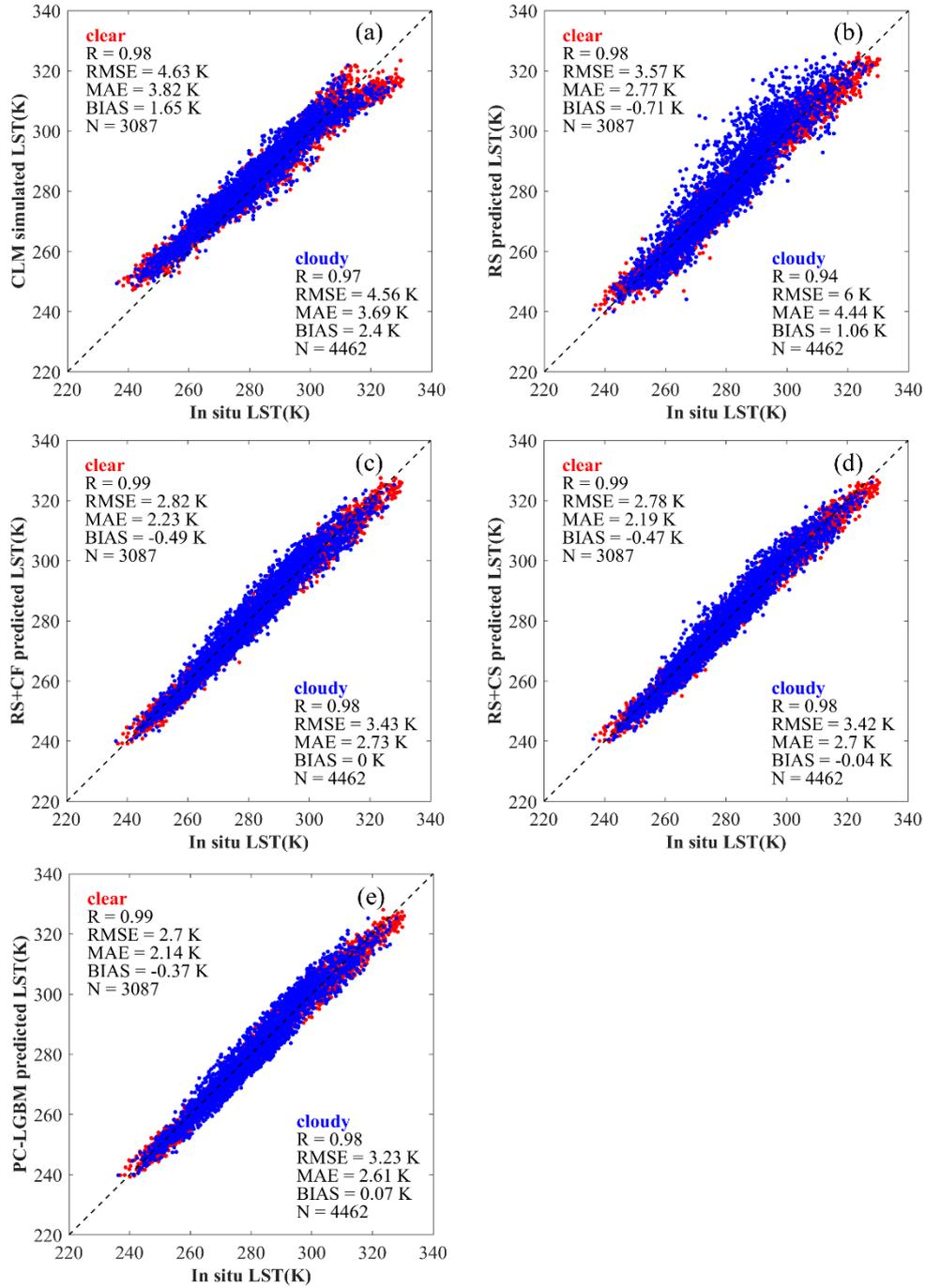

Fig. 8. Scatter plots of the all-weather LST against in-situ LST (four sites) during 2008 to 2011 with different data inputs: (a) CLM simulated; (b) RS; (c) RS+CF; (d) RS+CS; and (e) RS+CF+CS. The definitions of RS, CF and CS can be found in Table 3.

*4.3.2 Comparison with other advanced ML methods*

To further investigate the effectiveness of the proposed ML method, we selected three other advanced ML methods, including an XGBoost model (Chen and Guestrin 2016) and two deep learning models, i.e., a DBN (Hinton et al. 2006) and a GRNN (Specht 1991), as the comparative approaches, which are popular and widely used in climate and environment variable estimation (Li et al. 2020b; Shen et al. 2020; Tan et al. 2021b). It is worth noting that the random forest model was not used in this study, because it is time-consuming and inappropriate for tackling massive (e.g., 10 million level) datasets (Breiman 2001). Fig. 9 shows the model accuracy under the space-based validation for the different ML models using the same input data. As shown, all the ML methods achieve a reasonable accuracy, with an RMSE (R) of 2.62–3.55 K (0.97–0.99) during the daytime and 1.56–2.25 K (0.98–0.99) during the nighttime. The decision tree based boosting models (i.e., PC-LGBM and PC-XGBoost) perform better than the neural network based models (i.e., PC-DBN and PC-GRNN), which could be due to the advantages of boosting methods in countering overfitting problems when solving pixel-pixel regression tasks (Chen et al. 2022). In particular, the PC-LGBM model performs the best among all the physics-constrained ML methods, with an RMSE of 1.56–2.62 K and R of 0.99. Overall, the LGBM-based method outperforms the other popular ML methods in terms of model accuracy.

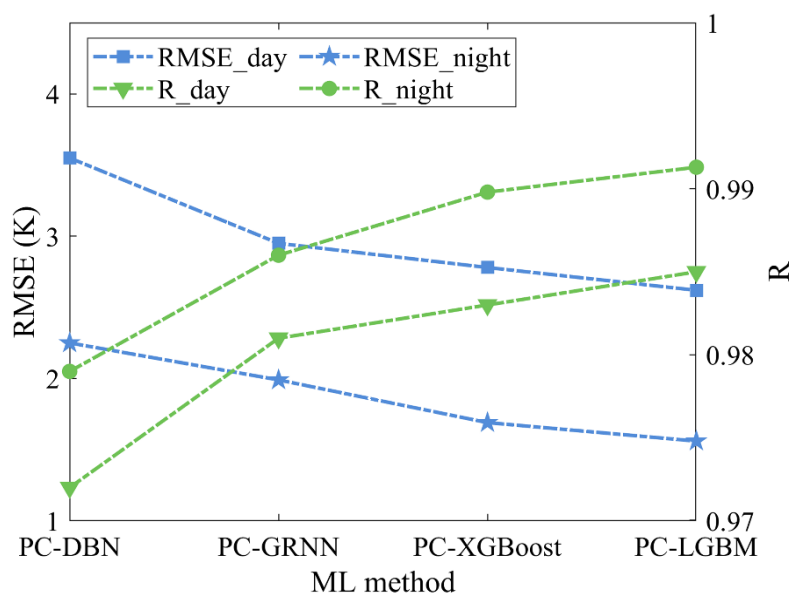

Fig. 9. Space-based model accuracy (RMSE and R) for the different ML models.

*4.3.3 Comparison with other all-weather LST data*

A number of fusion or reconstruction methods have been proposed to generate all-weather LST. To further evaluate the proposed method, we compared the PC-LGBM method with one interpolation method (Zhang et al. 2022) and three fusion-based methods (Wu et al. 2022; Xu and Cheng 2021; Zhang et al. 2021), which were recently proposed and proven to be effective in all-weather LST mapping. Note that the four afore-mentioned methods have been applied to produce gapless LST data or products covering the HRB-MU, so it is convenient to make such a comparison. To prevent any possible confusion, the products produced by Zhang et al. (2022) and Zhang et al. (2021) are referred to as the Zhang I's LST and Zhang II's LST, respectively. Fig. 10 shows a spatial comparison between the PC-LGBM LST and the four sets of all-weather LST data for DOY 106 in 2010, when the corresponding MODIS LST is almost all missing (not shown). It can be seen that all the LST data are spatially complete during both the daytime and nighttime, but show a difference in spatial patterns and LST magnitude. Obvious block effects and "blurred" effects are found in the Xu's LST (Fig. 10a). This phenomenon can be attributed to the coarse resolution of the AMSR2 data (i.e., $0.1° \times 0.1°$) used for multiresolution Kalman filtering in LST fusion (Xu and Cheng 2021). Since only spatial and temporal neighboring information of clear-sky LST is utilized in the reconstruction process proposed by Zhang et al. (2022), the reconstructed LST is a hypothetical "clear-sky" LST, resulting in a spatial difference between the Zhang I's LST and other LSTs in cloudy conditions. Some artifacts can be observed in the Wu's LST, because the original MODIS LST is retained. However, such a broken effect may not represent the real LST pattern due to the uncertainties in identifying the clear-sky pixels in MODIS LST (Wang et al. 2019). The uncertainties can lead to abnormal retrievals of partially cloud-contaminated pixels, particularly near the clustered cloud pixels (see Section 4.4) (Jia et al. 2022; Ma et al. 2020). Our LST shares a similar LST pattern with the Zhang II's LST, enhancing the spatial details and appearing more "natural", indicating the robustness of the proposed method in obtaining a more real LST.

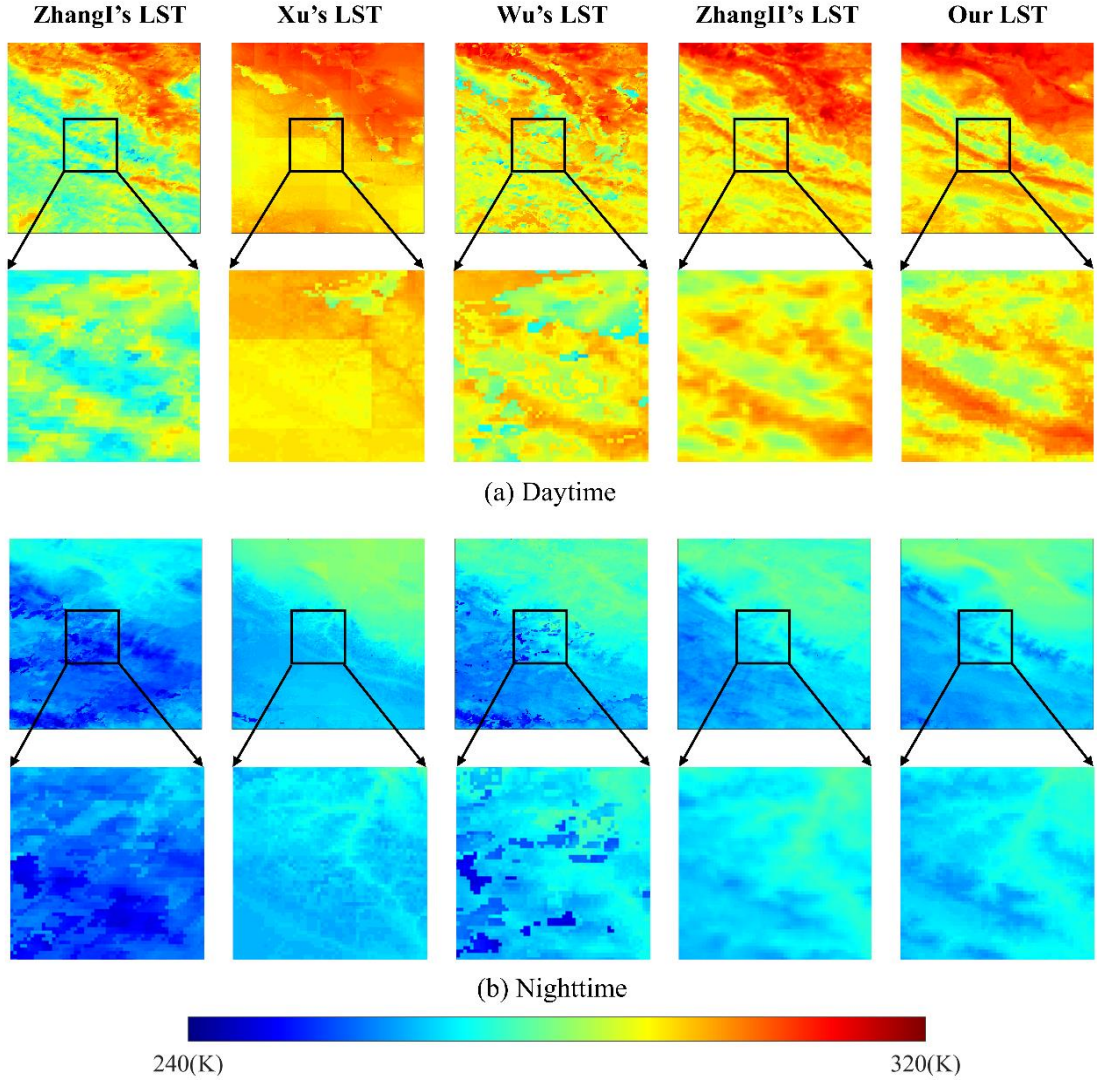

Fig. 10. Spatial patterns of the all-weather LST from the ZhangI's LST (Zhang et al. 2022), Xu's LST (Xu and Cheng 2021), Wu's LST (Wu et al. 2022), ZhangII's LST (Zhang et al. 2021) and our LST for DOY 106 in 2010 during (a) the daytime and (b) the nighttime. The black boxes denote certain regions for a detailed comparison.

To quantitatively compare the accuracy of the developed all-weather LST with the four sets of LST data, the same in-situ measurements from the AR, HZZ, and YK sites were used. The validation results under all-weather conditions during the daytime and nighttime are shown in Fig. 11. In general, all the gapless LST data have a relatively high accuracy, with R greater than 0.92 (0.97) and the RMSE less than 3.9 K (3.1 K) during the daytime (nighttime). Although the RMSE of our LST is slightly higher than that of some of the LST data (e.g., at the daytime AR site and the nighttime YK site), the mean RMSE remains the lowest, and the R of the PC-LGBM LST is higher than that of all the other LST data. The average values of R (RMSE) for the ZhangI's LST, Xu's LST, Wu's LST, ZhangII's LST and our LST from the three sites during both the daytime and nighttime are 0.96 (3.23 K), 0.96 (3.21 K), 0.96 (3.2 K), 0.96 (3.22 K), and 0.98 (3.01 K), respectively. Overall, the proposed method is superior to the above-mentioned fusion-based and reconstruction methods, based on both the mapping effect and validation accuracy over the HRB-MU.

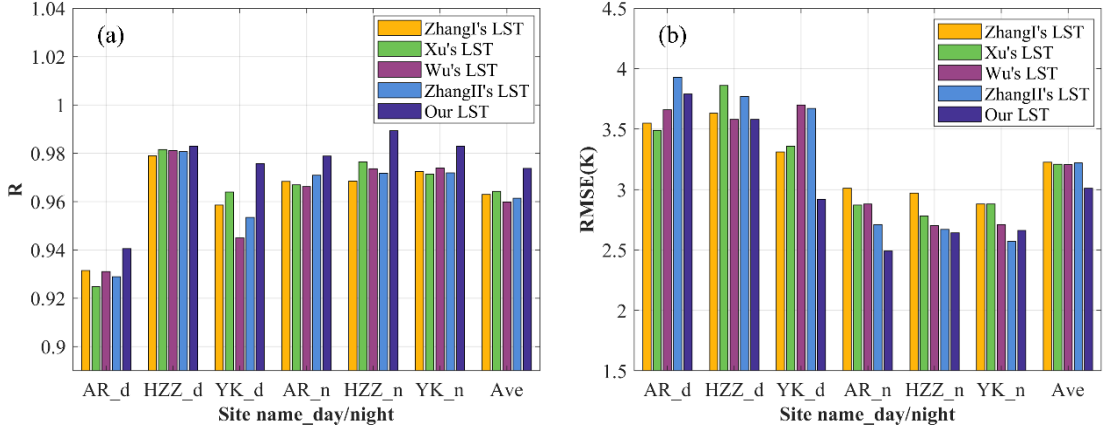

Fig. 11. Comparison between the all-weather PC-LGBM LST and the LST obtained by Zhang et al. (2022), Xu and Cheng (2021), Wu et al. (2022) and Zhang et al. (2021) against in-situ LST under all-weather conditions at the AR, HZZ, and YK sites in 2010: (a) R, (b) RMSE.

*4.4 Comparison between clear-sky LST and the original MODIS LST*

Clouds are prevalent over the land surface, and can be especially heavy in regions with complex terrain. To obtain more clear-sky MODIS LST samples for LST fusion (e.g., a similar pixel selection) or reconstruction, a number of studies have selected clear pixels with the "LST produced" and "average LST error <=3 K" according to the MODIS QC flags (hereafter termed "clear LST") (Duan et al. 2017; Gong et al. 2023; Long et al. 2019; Wu et al. 2022). However, owing to the solar-cloud-satellite geometry (SCSG) effect existing in the MODIS LST (Wang et al. 2019), LST detection based on poor QC flags may not represent the true situation. In this case, such a selection could weaken the accuracy of the fusion or reconstruction results.

In this study, the clear-sky MODIS LST pixels with the highest quality ("good quality" and "average LST error <= 1 K", hereafter termed "fully clear LST") were selected as the label data, and the likely-cloudy LST pixels ("other quality" and "average LST error > 1 K & <=3 K", hereafter termed "partially cloudy LST") were then recovered by PC-LGBM. Taking 2010 as an example, Fig. 12a shows the cumulative distribution plot of the spatially averaged percentages of the clear-sky samples for the daytime and nighttime over the HRB-MU. Overall, clear-sky LST pixels accounts for 55.3% over the HRB-MU, of which 38.4% are fully clear and a non-ignorable proportion of 16.9% are partially cloudless. Fig. 12b–c shows the scatter plots of the official MODIS LST under different conditions and the corresponding PC-LGBM LST against in-situ LST measurements in 2010. Under fully clear-sky conditions, the PC-LGBM LST is much closer to the MODIS LST, with an RMSE difference of 0.02 K, illustrating the excellent performance of the PC-LGBM model in reproducing the magnitude of clear-sky MODIS LST. Meanwhile, under partially cloudy conditions, the MODIS LST shows an unsatisfactory performance, followed by an RMSE of 5.31 K, MAE of 3.77 K, and a considerable negative BIAS of 2.58 K. It has been demonstrated that partially cloud-contaminated areas generally block the surface heating processes, leading to abnormally cool LSTs in the images (Jia et al. 2022). Correspondingly, the PC-LGBM LST largely makes up for this underestimation, with an RMSE of 3.72 K, MAE of 2.79 K, and a small BIAS of −0.45 K. In summary, the LST generated by the PC-LGBM model is comparable, or even better than the original MODIS LST under general clear-sky conditions.

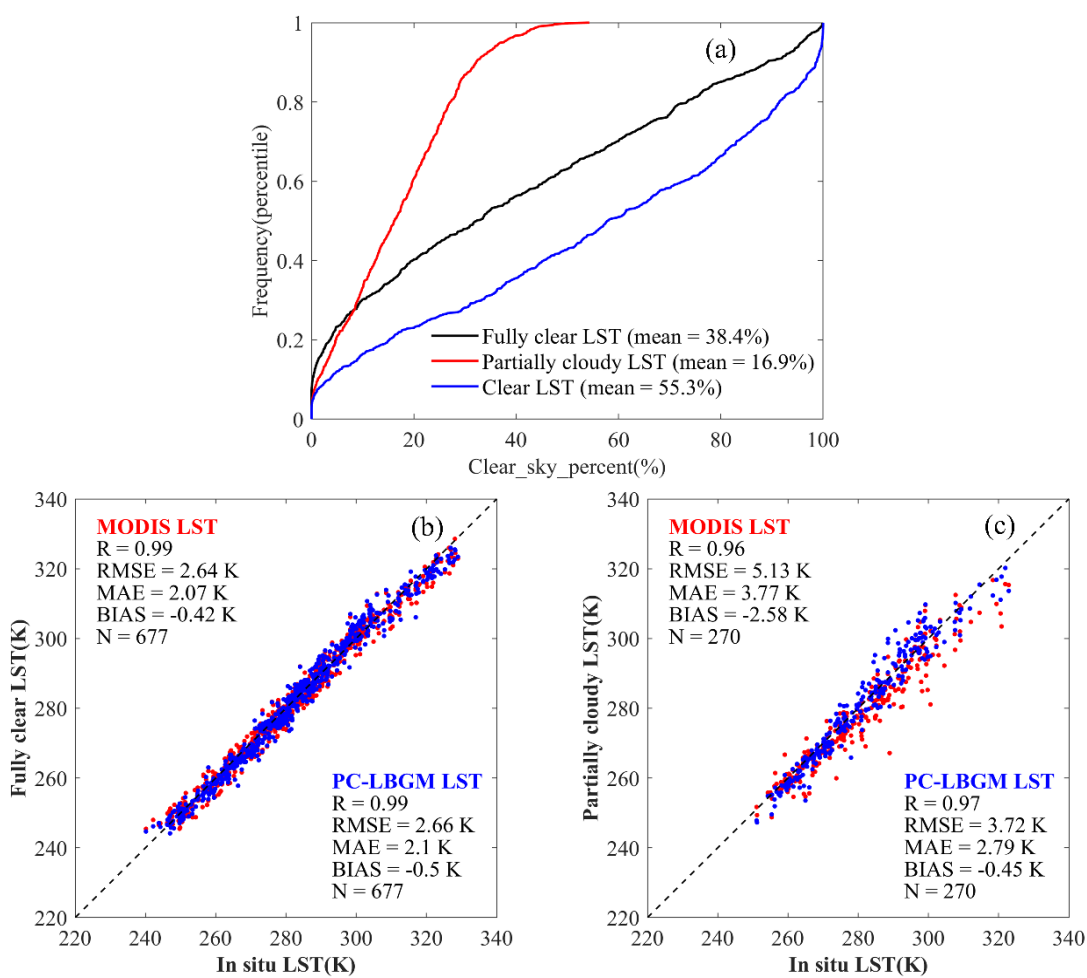

Fig. 12. (a) Cumulative distribution plots of the spatially averaged percentages of clear sky samples over the HRB-MU, and scatter plots of the official MODIS LST and PC-LGBM LST against in-situ LST measurements at the AR, HZZ, and YK sites for the daytime and nighttime in 2010 from (b) fully clear-sky pixels, and (c) partially cloud-covered pixels.

## 5 Discussion

*5.1 Effects of season, land-cover type and elevation on clear-sky LST estimation*

Previous studies have revealed the effect of the season on LST accuracy (Li et al. 2021; Wu et al. 2022). From this perspective, we divided all the training samples into eight subsets according to the season and daytime/nighttime. It is of special significance to evaluate the accuracies of the estimated LST in the different seasons. Table 4 lists the sample-based validation results in the different seasons during the daytime and nighttime. The performance of the PC-LGBM model varies in the different seasons, with a relatively low accuracy observed in the spring and summer during the daytime. The reason for this may be that the vegetation grows and peaks in these two seasons, which brings uncertainty to the LST estimation (Li et al. 2021). A relatively high accuracy is obtained in autumn, with the RMSE less than 1.8 K and R of 0.99. In addition, the season impact during the daytime is slightly higher than that during the nighttime. The difference between the maximum and minimum RMSE reaches 0.51 K and 0.38 K for the daytime and nighttime, respectively. Overall, the effect of season on the model accuracy exists, but is not significant, demonstrating the robust performance of the PC-LGBM model.

**Table 4**

Sample-based validation results in the different seasons for the daytime and nighttime, respectively. N denotes the sample size.

| Time | Season | RMSE (K) | MAE (K) | R | N |
| --- | --- | --- | --- | --- | --- |
| Daytime | spring | 2.3 | 1.71 | 0.98 | 4021766 |
| | summer | 2.1 | 1.53 | 0.98 | 3907006 |
| | autumn | 1.79 | 1.32 | 0.99 | 4840660 |
| | winter | 1.95 | 1.44 | 0.97 | 3633792 |
| Nighttime | spring | 1.25 | 0.93 | 0.99 | 6967453 |
| | summer | 1.03 | 0.73 | 0.99 | 6400269 |
| | autumn | 1.17 | 0.88 | 0.99 | 8191298 |
| | winter | 1.41 | 1.08 | 0.97 | 8097223 |

The heterogeneity of LST is significantly affected by different land-cover types (Holmes et al. 2009; Xu et al. 2022) and elevations (He et al. 2019; Zhao et al. 2019). Fig. 13 shows the performance of the PC-LGBM model by calculating the RMSE for different land-cover types and elevations. As shown in Fig. 13a, all the RMSE values are less than 2.27 K but varies across the different land-cover types, with the lowest value occurring in cropland and the highest value occurring in grassland during both the daytime and nighttime. The difference between the maximum and minimum RMSE for the PC-LGBM model is 0.56 K and 0.35 K for the daytime and nighttime, respectively. These results suggest that the proposed method demonstrates both a high accuracy and stability for different land-cover types.

To evaluate the impact of topography on the model performance, the model accuracies for different elevation ranges are presented in Fig. 13b. In general, as the elevation increases, the accuracy of the model tends to decrease, while relatively little change can be observed during the nighttime. However, there is a reverse trend when the elevation is above 3500–4500 m, which indicates that LST is comprehensively affected by different factors (Xu et al. 2022; Zhao and Duan 2020). Nonetheless, the RMSE values remain stable, at less than 2.4 K and 1.3 K during the daytime and nighttime, respectively. This indicates that the PC-LGBM model shows a stable accuracy in estimating clear-sky LST across different elevation ranges.

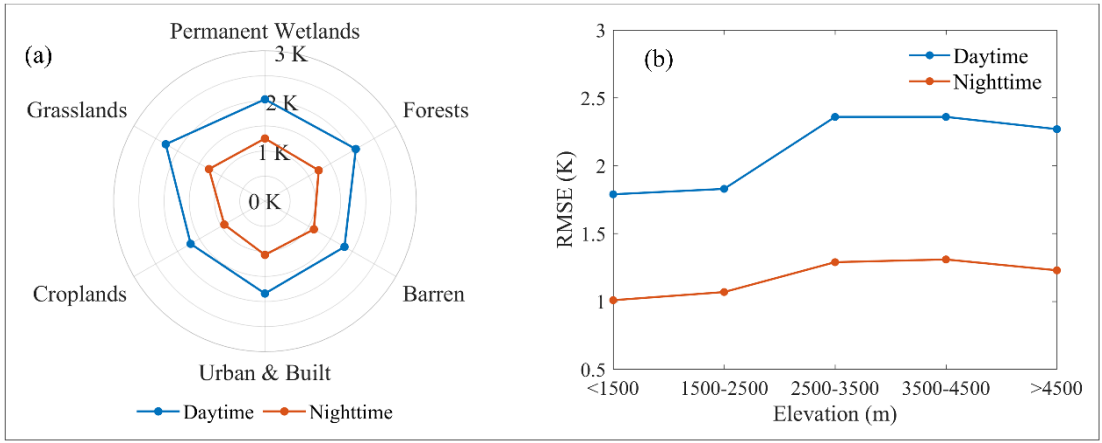

Fig. 13. Validation results (RMSE) of the PC-LGBM LST against MODIS LST in clear-sky conditions for specific (a) land-cover types and (b) elevation ranges.

*5.2 Variable importance analysis*

LST is a "fast" land surface variable that varies rapidly in space and time, which is affected by the surface conditions, topographic conditions, and atmospheric conditions (Crosson et al. 2012; Prata et al. 1995). In this study, we selected four sets of CLM meteorological forcing data (SRA, TMP, RHU, PRS), two sets of CLM simulation data (CLM-LST and CLM-SM), six sets of remote sensing data (DEM, LAT, NDVI, NDSI, B-VIS, B-NIR), and DOY, giving a total of 13 variables as the predictors. During the construction of the decision tree in the LGBM model, the collective benefits gained from the splits that utilize the feature of interest were used to measure the importance of the predictors. Fig. 14 displays the mean importance values of the predictors over the four seasons. The results reveal that CLM-LST and the four sets of meteorological forcing data are the main contributors to the model during both the daytime and nighttime, which confirms the SEB constraints theory in PC-LGBM. DOY and DEM also have important impacts on LST estimation, owing to their contributions in representing LST temporal variability and the terrain effect. In addition, SM is an essential factor contributing to the spatial heterogeneity of LST, due to its controlling impact on surface thermal inertia and evapotranspiration (Sandholt et al. 2002; Tang et al. 2010).

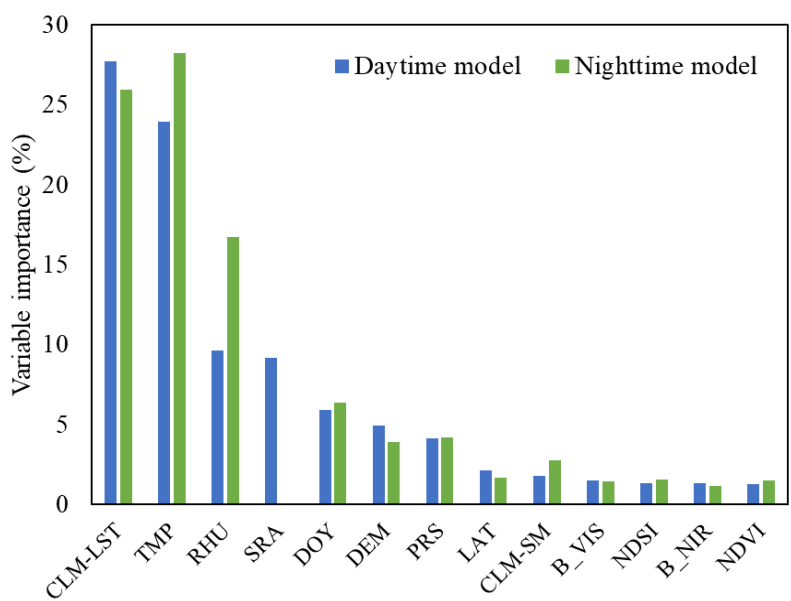

Fig. 14. Mean importance values of the input variables of the daytime model and nighttime model.

*5.3 Advantages and limitations of the proposed method*

The advantages of the proposed method are reflected in the following aspects. Firstly, the previous studies usually applied the relationship of the clear-sky model to restore cloudy LST, without sufficient consideration of the cloud effects on LST (Buo et al. 2021; Xiao et al. 2021). Meanwhile, the linking predictors were typically temporally sparse (e.g., 3-h to 16-day), which resulted in uncertainty in presenting the high temporal variability of LST (Cho et al. 2022; Li et al. 2021). In response to this issue, the PC-LGBM model fully maximizes the potential of proxy data for clouds with a high spatio-temporal resolution (e.g., reanalysis data and model-simulated data with ~7 km and 1-h resolutions), which can discriminate between clear-sky and cloudy-sky conditions, resulting in a good and comparable accuracy in the estimated gapless LST under both clear-sky and cloudy-sky conditions (see Section 4.2).

Secondly, it is acknowledged that ML models suffer from "black box" problems, which has led to an absence of physical meaning in the reconstrued cloudy LST. Inspired by the advances in coupling process understanding and ML in ESS (Reichstein et al. 2019), we investigated the rationality and feasibility of incorporating explicit physical constraints into an implicit ML model to improve LST prediction accuracy and physical interpretability. Special attention was paid to the evaluation of uncertainty in different data combinations and the effects of physical intervention. Compared with the pure ML models (e.g., RS, RS+CF, and RS+CS in Section 4.3.1), the PC-LGBM model (i.e., RS+CF+CS) achieved a superior performance. It was demonstrated that the underlying process knowledge (e.g., the SEB and the law of heat conduction, see Eq. (12)) between the key LSM forcing data and LSM simulation data in the CLM was mined and applied in the learning process, so that the learning model has relatively high interpretability. Meanwhile, ML models are known to have extrapolation problems, and coupling such physical constraints helped to capture the LST values under extreme weather conditions, thereby enhancing the generalization ability of the ML model (see Section 4.1).

Thirdly, benefiting from the utilization of the highest-quality MODIS LST and the powerful data mining ability of ML, the likely cloudy MODIS LST was corrected. Theoretically and practically, the LST obtained in this study achieved a comparable or even better accuracy than the original MODIS LST in general clear-sky conditions (see Section 4.4). Lastly, the proposed method is currently applicable to the HRB-MU with CLDAS data. Given that all the remote sensing data used are gapless and worldwide, and provided that the LSM forcing data and simulation data are updated on a larger scale, or replaced with other global-scale data (e.g., GLDAS and European Centre for Medium-Range Weather Forecasts Reanalysis v5 (ERA5)), this method could be easily expanded to other regions. Moreover, the PC-LGBM model has excellent efficiency, so once the model is trained, it could be readily applied for generating gapless and long time-series LST products.

Nevertheless, there are also some uncertainties and limitations to this method. We compared our method with other excellent LST reconstruction methods, demonstrating a superior performance of the PC-LGBM model (see Section 4.3.3). However, this comparison was conducted at a regional scale, and further investigation is needed for a larger-scale comparison. Another uncertainty is the spatial representativeness of the ground sites. Although previous studies have evaluated the spatial representativeness of the sites in the WATER observatory network and demonstrated that the sites used in the study were qualified for homogeneity validation (Li et al. 2009; Yu et al. 2014a), the scaling effects induced by spatial mismatch cannot be totally neglected, which is a typical difficulty in remote sensing product validation (Duan et al. 2019; Li et al. 2020a). In this study, LSM-related data were used as a proxy to reflect cloud effects. However, the true "cloudy" label data were still unavailable. Further studies are advocated to incorporate additional ground-based measurements or PMW information into the learning-based model. Lastly, we mainly verified the effectiveness of coupling the input variable physical constraints into the ML model in this study (Shen and Zhang 2023). In future work, more attention should be paid to further investigating the different constraint construction methods, such as objective function constraints (e.g., imposing model penalties in the cost function for violating the SEB) and model structure constraints (e.g., adding a physical relationship expressing the SEB to the middle or back of the neural network).

## 6 Conclusion

Cloud contamination has hindered the access to gapless LST for potential applications. In this paper, we have proposed a physics-constrained ML model (PC-LGBM) by integrating key CLM forcing data, CLM simulation data and remote sensing data into an ML model. The PC-LGBM model combines the advantages of process knowledge (SEB physical constraints, interpretability) and ML (data adaptability). The well-trained model was applied to instantaneous observations from Aqua MODIS to generate a daily $0.01° \times 0.01°$ gapless LST dataset from 2008 to 2011 over the HRB-MU. The generated LST showed excellent spatio-temporal continuity and retained high consistency with the MODIS LST in spatial details. In terms of temporal variability, the generated LST can basically capture the seasonal and daily changes of LST. In addition, the PC-LGBM model showed a robust performance across different seasons, land-cover types, and elevation ranges. The accuracy of the clear-sky LST generated by the PC-LGBM model was similar to that of the original MODIS LST, or even superior in certain scenarios. The reconstructed cloudy LST against in-situ measurements showed a satisfactory accuracy (RMSE = 2.91–3.66 K, MAE = 2.35–3 K, R = 0.97–0.98), which was generally comparable to that obtained under clear-sky conditions. Among the four popular ML methods, the LGBM-based model performs the best in model accuracy. Compared with a pure physical method and pure ML methods, the PC-LGBM model improved the LST prediction accuracy, physical interpretability, and generalization ability of the LST estimation, especially for extreme weather cases. When compared with other all-weather LST data, it was found that the PC-LGBM LST was more "real" and "natural" and had a higher accuracy over the HRB-MU.

**Credit author statement**

**Declaration of competing interest**

The authors declare no conflict of interest.

**Acknowledgements**

This study was jointly supported by the Key Projects of the National Natural Science Foundation of China under grant 42130108, the Special Fund of Hubei Luojia Laboratory under grant 220100041, and the National Natural Science Foundation of China under grant 42271381. The anonymous reviewers' and editors' comments were highly appreciated. The MODIS products were accessed from the Earthdata website (https://search.earthdata.nasa.gov/search). The GLASS products were downloaded from http://www.glass.umd.edu/. The soil data were downloaded at http://globalchange.bnu.edu.cn/. The in-situ data were provided by the National Tibetan Plateau Data Center (http://data.tpdc.ac.cn). The CLDAS forcing data were provided by the China Meteorological Administration (http://tipex.data.cma.cn/tipex). The Zhang I's LST was obtained from https://doi.org/10.25380/iastate.c.5078492. The Zhang II's LST was obtained from https://data.tpdc.ac.cn. The Xu's LST was obtained from http://www.geodata.cn. The Wu's LST was provided by Wu et al. (2022). The numerical calculations in this paper have been done on the supercomputing system in the Supercomputing Center of Wuhan University.